\documentclass[10pt]{iopart}
\usepackage{CJK}
\usepackage{cite}
\usepackage{ulem}
\usepackage{bbm,amsmath,amssymb}
\usepackage{graphicx}
\usepackage{dcolumn}
\usepackage{xcolor}
\usepackage{bbold}
\usepackage{tikz}
\usepackage{blkarray}

\newcommand\redsout{\bgroup\markoverwith{\textcolor{red}{\rule[0.5ex]{2pt}{0.4pt}}}\ULon}

\begin{document}
\begin{CJK*}{GBK}{ } 
\title[]{The effects of gravitational waves on a hydrogen atom}

\author{N Wanwieng$^1$, N Chattrapiban$^1$ and A Watcharangkool$^{2}$}

\address{$^1$Department of Physics and Materials Science, Faculty of Science, Chiang Mai University, Chiang Mai 50200, Thailand}
\address{$^2$ National Astronomical Research Institute of Thailand, Chiang Mai 50180, Thailand}
\eads{\mailto{nontapat\_w@cmu.ac.th}, \mailto{apimook@narit.or.th}}
\vspace{10pt}
\begin{indented}
\item[]August 2023
\end{indented}

\begin{abstract}
We investigate the influence of gravitational waves on a freely falling hydrogen atom by analyzing the dynamics of the bound electron described by the Dirac equation in the curved spacetime of a gravitational wave. From this, we derive the corresponding Dirac Hamiltonian in the Local Inertial Frame of the atom, assuming gravitational waves are described by the linearized theory of General Relativity. To maintain meaningful physical interpretations while obtaining a non-relativistic description, we employ the Foldy-Wouthuysen transformation. Through the analysis of resulting interaction terms and comparison with flat spacetime counterparts, valuable insights into the effects of gravitational waves on the hydrogen atom are gained. Additionally, we explore selection rules governing the coupling between gravitational waves and the atom and utilize first-order perturbation theory to quantify the induced energy shifts and spectral line splitting. This investigation contributes to our understanding of the interplay between quantum systems and gravitational waves, which could lead to alternative method of gravitational waves indirect detection. However, measuring such tiny energy shifts would require a telescope with very high spectral resolution.
\end{abstract}

%
\vspace{2pc}

\ioptwocol

\section{Introduction}
1
The groundbreaking discovery of gravitational waves (GWs) \cite{abbott2016} has revolutionized astronomy and cosmology, providing a new avenue to explore the cosmos and probe the fundamental nature of gravity \cite{bailes2021,mastrogiovanni2022}. Despite these remarkable advancements, the influence of GWs on the quantum behavior of fundamental particles remains an intriguing open question. Understanding this interaction might help us develop a clearer path toward a robust theoretical framework for quantum gravity. While a complete theory of quantum gravity remains elusive, significant progress can be made through a semi-classical approach.

Over the years, several researchers have utilized the Dirac equation to gain valuable insights into the interplay between gravity and quantum effects \cite{parker1980, Parker1981,Pimentel1982,Gill1987,obukhov2001,Marques2002,Zhao2007,obukhov2011,quach2015,KOKE2016}. Parker and colleagues \cite{parker1980,Pimentel1982} explored the potential of using the one-electron atom as a probe to investigate spacetime curvature, pioneering a possibility of characterizing spacetime through atomic systems. Gill et al. \cite{Gill1987} delved into the behavior of hydrogen-like atoms in strong gravitational fields, offering insights into the quantum behavior of particles under extreme gravitational conditions. Obukhov's work \cite{obukhov2001} examined the intriguing connection between spin, gravity, and inertia for Dirac particles, offering valuable insights into their coupling. Also, in 2011, Obukhov et al. explored the behavior of Dirac fermions in strong gravitational fields in a different context \cite{obukhov2011}. Marques and Bezerra \cite{Marques2002} extended the analysis to study the behavior of hydrogen atoms in the presence of topological defects in the gravitational field, uncovering the effects of curvature singularities and defects on the atom's dynamics. Zhao et al. \cite{Zhao2007} focused on the energy shifts of a stationary hydrogen atom in a static external gravitational field with Schwarzschild geometry. Shwartz and Giulini \cite{Schwartz2019} presented a precise post-Newtonian Hamiltonian description of an atom in a weak gravitational field, providing a valuable framework for describing atomic systems influenced by weak gravitational fields.

In addition to theoretical investigations, experimental research has significantly enhanced our understanding of quantum phenomena in gravitational fields. In 2002, Nesvizhevsky et al. \cite{nes2002} provided pioneering evidence of the wave-like behavior of neutrons in the presence of gravity, marking the inception of experiments exploring quantum effects within gravitational contexts. Subsequently, Ichikawa et al. \cite{Ichikawa} conducted experiments on antimatter, emphasizing the significance of testing matter-antimatter symmetry in gravitational fields. In 2020, Asenbaum \cite{asenbaum2020} demonstrated the precision capabilities of cold atom quantum sensors in detecting gravitational effects, offering new avenues for high-precision experiments. Most recently, Overstreet et al. \cite{overstreet2022} unveiled quantum behavior in macroscopic objects, pushing the quantum-classical boundary within gravitational systems.

Despite the theoretical interest in the effects of gravitational fields on atoms, the investigation of GW effects on atoms, especially in the non-relativistic regime, has received limited attention. To bridge this gap, we provide a comprehensive analysis of the non-relativistic limit of the Dirac equation in a GW background. The paper's objectives are three-fold: (1) to derive the Dirac Hamiltonian describing the dynamics of a hydrogen atom in the presence of GWs; (2) to explore the selection rules governing the coupling between GWs and the atom; and (3) to quantify the energy shifts and spectral line splitting induced by GWs.

To achieve these objectives, GWs are treated within the framework of the linearized theory of General Relativity, focusing on the primary effects while ensuring mathematical tractability. The non-relativistic limit is obtained through the modified Foldy-Wouthuysen transformation developed by M. Buhl et al. \cite{approxFW}, enabling the construction of a non-relativistic theory with relativistic corrections that preserves meaningful physical interpretations. By analyzing the resulting interaction terms and comparing them with those in flat spacetime, we gain valuable insights into the effects of GWs on the dynamics of the hydrogen atom.

The organization of this paper is as follows: Section \ref{secII} introduces the Dirac equation for an electron bound by the Coulomb interaction with the nucleus in the presence of a linearized mono-frequency GW, deriving the corresponding Dirac Hamiltonian. In Section \ref{secIII}, we apply the modified Foldy-Wouthuysen transformation to obtain a new representation suitable for studying the non-relativistic limit, resulting in the non-relativistic Hamiltonian with interaction terms. Section \ref{secIV} investigates the perturbative effect of the linearized GW on a hydrogen atom, calculating the induced energy shifts resulting from the dominant GW interactions. Finally, in Section \ref{secV}, we discuss the magnitude of these energy shifts and their implications. Throughout the paper, spacetime indices of tensors are denoted by Greek letters. Additionally, we adopt a metric signature of $(+ - - -)$.

\section{Dirac equation in the linearized gravitational wave background}
\label{secII}

To investigate the effects of GWs on a freely falling hydrogen atom in space, we establish a local inertial frame (LIF) that follows the atom's geodesics. This frame is described using Fermi-normal coordinates, with the origin located at a point along the proton's geodesic. In this coordinate system, the time-like coordinate is represented as $x^0 = ct$, which corresponds to the proper time measured by an ideal clock carried by the atom. The spatial coordinates $x^i$ represent the proper distance measured by an ideal ruler carried by the atom near the geodesic, along orthogonal axes. With this coordinate system, the spacetime metric in the vicinity of the origin accurate to linear order in Riemann tensor reads
\begin{eqnarray}\label{metricFermi}
ds^2 \simeq \; c^2dt^2 \left(1 + R_{i0j0} x^i x^j \right)\notag\\
- dx^i dx^j \left(\delta ^{ij}
- \frac{1}{3}{R_{ikjl}} x^k x^l \right) \notag\\
+ 2cdtdx^i \left( \frac{2}{3} R_{0jik} x^j x^k \right), 
\end{eqnarray}
where the Riemann tensor is evaluated at the origin, and the Latin indices represent spatial components. Notably, the effect of spacetime curvature is manifested at the quadratic order of $x^i$.

In the context of the linearized theory of GWs, we can express the Riemann tensor in the LIF in terms of the transverse-traceless (TT) components of the metric perturbation, denoted as $h_{\mu\nu}^{\mathrm{TT}}$. The expression is given by
\begin{eqnarray}\label{Rieman}
    R_{\alpha \beta \gamma \delta } = \frac{1}{2} (h_{\alpha \delta ,\beta \gamma }^{\mathrm{TT}} + h_{\beta \gamma ,\alpha \delta}^{\mathrm{TT}} - h_{\alpha \gamma ,\beta \delta }^{\mathrm{TT}} - h_{\beta \delta ,\alpha \gamma }^{\mathrm{TT}} ),
\end{eqnarray}
where the comma denotes the partial derivative, e.g. $h_{\alpha \delta ,\beta \gamma}^{\mathrm{TT}} = \partial^2 h_{\alpha \delta }^{\mathrm{TT}}/\partial x^\beta \partial x^\gamma$.

For simplicity, let's consider the propagation of a plus-polarized GW along the z-axis of the LIF. In this scenario, the only non-zero components of the transverse-traceless metric perturbation are
\begin{eqnarray}
    h_{xx}^{\mathrm{TT}} = - h_{yy}^{\mathrm{TT}} = h_+(t), 
\end{eqnarray}
where $h_+(t)$ describes the waveform of the wave. As a result, the only non-zero component in \eqref{Rieman} is $R_{0i0j} = -(1/2c^2) \ddot h ^{\mathrm{TT}}_{ij}$, where the dots represent the time derivatives. The spacetime metric, \eqref{metricFermi}, reads
 \begin{eqnarray}\label{metricGWFermi}
   ds^2 \simeq \; \left(1 - \frac{1}{2c^2}\ddot h_{lm}^{\mathrm{TT}}x^l x^m\right)c^2 dt^2 - \delta _{ij} dx^i dx^j,
\end{eqnarray}
where
\begin{eqnarray}\label{metriccom}
    g_{00} = 1 - \frac{1}{2c^2}\ddot h_{lm}^{{\mathrm{TT}}} x^l x^m,\; g_{ij} =  - \delta _{ij},\; g_{0i} = 0,
\end{eqnarray}
and, to first order in $|h^{\mathrm{TT}}_{\mu\nu}|$, the inverse metric tensors are
\begin{eqnarray}\label{inversemetriccom}
    g^{00} = 1 + \frac{1}{2c^2}\ddot h_{lm}^{\mathrm{TT}} x^l x^m,\; g^{ij} =  - \delta ^{ij},\; g^{0i} = 0.
\end{eqnarray}
The behavior of an electron within a hydrogen atom, under the influence of a GW, can be described using the general covariant Dirac equation:
\begin{eqnarray}\label{Diraceq}
\left[ i\hbar {\bar \gamma }^\mu \left( D_\mu  + \frac{iq}{\hbar} A_\mu  \right) - mc \right]\Psi  = 0,
\end{eqnarray}
(see \cite{collas2019} for more details.) where $\Psi$ is the Dirac spinor, ${\bar \gamma}^\mu (x)$ represents the spacetime-dependent gamma matrices at a point $x$, and is defined by the Clifford algebra:
\begin{eqnarray}
    {\bar \gamma }^\mu (x) {\bar \gamma }^\nu (x) + {\bar \gamma }^\nu (x) {\bar \gamma }^\mu(x) = 2g^{\mu\nu }(x).
\end{eqnarray}
Here, $g_{\mu\nu}$ is the spacetime metric. The bared-gamma matrices are related locally to the constant (unbared) gamma matrices through the tetrad and its inverse at that point:
\begin{eqnarray}
    {\bar{\gamma}}^\mu (x) = {e_{\hat{a}}}^\mu (x) \gamma ^a, \quad \gamma ^a = {e^{\hat{a}}}_\mu (x) {\bar{\gamma}}^\mu (x).
\end{eqnarray}
The tetrads ${e_a}^\mu$ and their inverses ${e^a}_\mu$ are defined to satisfy the relations ${e^a}_\mu {e_b}^\mu = \delta^a_b$ and ${e^a}_\mu {e_a}^\nu = \delta^\mu_\nu$.

In this paper, we adopt the standard representation of the Dirac gamma matrices:
\begin{eqnarray}
    \gamma ^0 = \left( \begin{array}{cc}
\mathbb{1}_2 & 0\\
0 & -\mathbb{1}_2
\end{array}\right),\quad \gamma ^k = \left(\begin{array}{cc}
0&\sigma^k\\
 - \sigma^k&0
\end{array} \right),
\end{eqnarray}
where $\sigma ^k$ denotes the Pauli matrices. In this representation, $\gamma ^0$ is Hermitian, while $\gamma ^k$ is anti-Hermitian, satisfying $(\gamma ^0)^\dag = \gamma^0$ and $(\gamma ^k)^\dag = - \gamma ^k$. Since we are working with the signature $-2$, i.e., $\eta _{ab} = \mathrm{diag}(+,-,-,-)$, we have $\gamma _0 = \gamma ^0$ and $\gamma _k = - \gamma ^k$. 
The spinorial covariant derivative $D_\mu = \partial_\mu + \Gamma_\mu$ incorporates the spin connection $\Gamma_\mu = (1/8)\omega_{ab\mu} [\gamma^a,\gamma^b]$, where $\omega_{ab\mu} = g_{\rho\sigma}{e_{\hat a}}^\rho \nabla_\mu {e_{\hat b}}^\sigma$. Given a $3+1$ splitting of spacetime, the Dirac equation can be reformulated in terms of the Schr\"{o}dinger equation,
\begin{eqnarray}
 i\hbar\frac{\partial}{\partial t}\Psi  = H\Psi,  
\end{eqnarray}
where the corresponding Dirac Hamiltonian is given by
\begin{eqnarray}\label{HD}
    H =  &- i\hbar c\frac{ {\bar \gamma }^0 {\bar \gamma }^k}{g^{00}}\left(\partial _k + i\frac{q}{\hbar } A_k + \Gamma _k \right) - i\hbar \Gamma _0 + qc A_0 \notag\\
    &+ \frac{{\bar \gamma }^0}{g^{00}}mc^2 - \frac{i}{2}\frac{{\bar \gamma }^0 {\gamma ^0}}{g^{00} \sqrt { - g}} \frac{\partial }{\partial t}\left( \sqrt { - g} {\gamma ^0}{\bar \gamma }^0 \right).
\end{eqnarray}
The last term is included to ensure Hermiticity of the Hamiltonian with respect to the preserved scalar product in a generic curved spacetime, as introduced and discussed in Huang et al. \cite{huang2009}.

The 4-potential that describes the Coulomb interaction between the electron and nucleus in the presence of a linearized GW can be obtained by solving Maxwell's equations in the curved spacetime described by \eqref{metricGWFermi}, while considering appropriate boundary conditions. The solutions for the components of the 4-potential, accurate to linear order in $h_+$ and using Gaussian units, are given by
\begin{eqnarray}
    A^0 = \frac{e}{r}\left(1 - \frac{1}{{8c^2}}\ddot h_{lm}^{\mathrm{TT}}x^lx^m \right),\quad A^k = 0,
\end{eqnarray}
where $e$ represents the elementary charge and $r$ denotes the radial distance between the electron and nucleus.

In order to construct the Dirac Hamiltonian for the atom in the spacetime background, we find it advantageous to adopt a canonical Minkowski tetrad, which is defined by
\begin{eqnarray}\label{cantetrad}
    {e_{\hat k}}^0 = 0, \quad {e_{\hat 0}}^0 = \sqrt{ - g^{00}}. 
\end{eqnarray}
The rationale behind this choice will be clarified when we transition to the non-relativistic limit in Section \ref{secIII}.

The corresponding inverse tetrads are
\begin{eqnarray}\label{GWtetrad}
    {e_{\hat 0}}^0 = 1 + \frac{1}{4c^2}\ddot h_{lm}^{\mathrm{TT}}{x^l}{x^m},\quad {e_{\hat k}}^i = \delta _k^i,\quad {e_{\hat 0}}^i = {e_{\hat i}}^0 = 0.
\end{eqnarray}

Substituting \eqref{metriccom}, \eqref{inversemetriccom}, and \eqref{GWtetrad} into \eqref{HD}, the Dirac Hamiltonian \eqref{HD} becomes
\begin{eqnarray}\label{HGW}
    H = H_0 + H_I
\end{eqnarray}
with
\begin{eqnarray}\label{H0}
    H_0 = \beta mc^2 - i\hbar c \alpha ^k \partial _k - \frac{e^2}{r}
\end{eqnarray}
and 
\begin{eqnarray}\label{HI}
H_I =&  - \frac{1}{4c^2}\ddot h_{lm}^{\mathrm{TT}} x^l x^m \beta mc^2 + \frac{i\hbar c}{4}\ddot h_{lm}^{\mathrm{TT}} x^l  x^m \alpha ^k \partial _k \notag\\
&+ \frac{i\hbar c}{4}\ddot h_{jk}^{\mathrm{TT}} x^j \alpha ^k + \frac{1}{8}\frac{e^2}{r}\ddot h_{lm}^{\mathrm{TT}} x^l x^m.
\end{eqnarray}
where $\beta \equiv \gamma^0$ and $\alpha^k \equiv \gamma^0 \gamma^k$.
\section{The Hamiltonian in non-relativistic limit}\label{secIII}
The Foldy-Wouthuysen transformation (FWT) is a powerful technique employed to extract the non-relativistic limit of the Dirac equation while retaining the essential relativistic features. We apply this method to the Dirac equation in the presence of GWs. 

Originally, the FWT was specifically developed to analyze weak-field electromagnetic interactions \cite{FW}. By employing a successive unitary transformations parameterised as $e^{iS}$, where $S$ represents a hermitian operator. The Dirac Hamiltonian and the spinor wave function transforms according to
\begin{eqnarray}\label{Hexpand}
&H' = e^{iS(t)}H e^{ - iS(t)} - e^{iS(t)}\dot S e^{ - iS(t)} \notag\\
&= H + i[S,H] + \frac{i^2}{2!}[S,[S,H]] + \frac{i^3}{3!}[S,[S,[S,H]]] + \ldots \notag\\
 &- \dot S - i[S,\dot S] - \frac{i^2}{2!}[S,[S,\dot S]] + \frac{i^3}{3!}[S,[S,[S,\dot S]]] + \ldots
\end{eqnarray}
and
\begin{eqnarray}
\Psi' = e^{iS(t)}\Psi.
\end{eqnarray}

To proceed, we decompose the Dirac Hamiltonian as $H = \beta mc^2 + \mathcal{O} + \mathcal{E}$, with $\mathcal{O}$ and $\mathcal{E}$ denoting the odd and even parts of $H$, respectively. The odd part $\mathcal{O}$ is defined by the anticommutator $\{\mathcal{O}, \beta\} = 0$, while the even part $\mathcal{E}$ is defined by the commutator $[\mathcal{E}, \beta] = 0$. The success of the FWT hinges upon a judicious choice of the operator $S$ that effectively eliminates the odd part, up to the desired order. For weak-field electromagnetic interactions, the FWT is achieved by choosing $S = i\beta \mathcal{O}/mc^2$. However, when dealing with the Dirac Hamiltonian in a curved spacetime, challenges arise due to the presence of terms of order $mc^2$, making the application of the FWT more intricate. This limitation stems from the Weak Equivalence Principle, which dictates the equality of inertial mass and gravitational mass---a distinction absent in the electromagnetic case. Consequently, the original FWT fails to fully account for the effects of the gravitational field on the Dirac equation. Addressing these challenges requires a more refined approach to incorporate the gravitational effects accurately.

To overcome these limitations, we adopt the method developed by M. Buhl et al. \cite{approxFW}. This approach addresses the challenges posed by the presence of terms of order $mc^2$ in the Dirac Hamiltonian and aims to reduce the odd part describing the interaction with the gravitational field. To achieve this, it is essential to constrain the tetrad frame to a canonical Minkowski form, as defined in \eqref{cantetrad}.

In the canonical Minkowski tetrad, the Dirac Hamiltonian takes the general form 
\begin{eqnarray}\label{Hgen}
    H = \beta mc^2 + \beta Amc^2 + B_k \sigma ^k \mathbb{1}_4 + D\mathbb{1}_4 + \notag\\
    {F_k}{\alpha ^k} + {G_k}^i{\alpha ^k}{\partial _i} + J{\gamma ^5},
\end{eqnarray}
where the coefficients depend on the tetrads and metric, and they can be computed directly using the formulas given in \cite{approxFW}. To implement the FWT, we use the appropriate operator given by
\begin{eqnarray}\label{S}
    S = \frac{i\sqrt{g^{00}}}{2mc^2}(K_a \gamma ^a + {G_a}^i{\gamma ^a}{\partial _i} + J\beta {\gamma ^5})
\end{eqnarray}
with
\begin{eqnarray}\label{Ka}
    {K_a} = {F_a} - \frac{{i\hbar c}}{{4{{({g^{00}})}^{3/2}}}}({\partial _i}{g^{00}}){e_{\hat a}}^i.
\end{eqnarray}
In our case, it is found that $K_a\gamma^a=0$. By comparing \eqref{HGW} to  \eqref{Hgen}, we can determine the coefficients as follows
\begin{eqnarray}
A &=  - \frac{1}{4c^2}\ddot h_{lm}^{{\mathrm{TT}}}{x^l}{x^m}\\
B_k &= J = 0\\
D &=  - \frac{{{e^2}}}{r}(1 - \frac{1}{8c^2}\ddot h_{lm}^{{\mathrm{TT}}}{x^l}{x^m})\\
F_k &=   \frac{{i\hbar }}{4c}\ddot h_{jk}^{\mathrm{TT}}{x^j} \label{Fa}\\
{G_k}^i &=  - i\hbar c(1 - \frac{1}{4c^2}\ddot h_{lm}^{{\mathrm{TT}}}{x^l}{x^m})\delta _k^i.
\end{eqnarray}
By substituting \eqref{Fa} into \eqref{Ka}, along with \eqref{GWtetrad}, we obtain
\begin{eqnarray}
    K_a =   \frac{i\hbar}{4c}\ddot h_{ja}^{\mathrm{TT}}{x^j} - \frac{{i\hbar }}{4c}\ddot h_{im}^{{\mathrm{TT}}}{x^m}{e_a}^i.
\end{eqnarray}
Subsequently, a direct calculation yields $K_a\gamma^a$ = 0.
Therefore, the operator $S$ in  \eqref{S} takes the form
\begin{eqnarray}\label{SFW}
   S =  - \frac{i}{2mc}\beta {\alpha ^k}{p_k},
\end{eqnarray}
where $p_k = -i\hbar \partial _k$ is the momentum operator. Notably, the operator $S$ in  \eqref{SFW} does not explicitly depend on time and is proportional to $1/mc^2$. When we expand the transformed Hamiltonian, \eqref{HGW}, as a power series of $1/mc^2$, we restrict ourselves to considering only terms up to order $1/mc^2$. This enables us to express the transformed Hamiltonian as
\begin{eqnarray}\label{HH}
    H_{\mathrm{FW}} =& H + i[S,H] + \frac{i^2}{2}[S,[S,\beta mc^2(1 - \frac{1}{4c^2}\ddot h_{lm}^{\mathrm{TT}} x^lx^m)]] \notag\\
    &+ O\left( \frac{1}{m^2c^4} \right).
\end{eqnarray}
By computing the various commutators in \eqref{HH} (See Appendix A), and neglecting the odd terms, which are of order $(1/mc^2)^2$, we then have 
\begin{eqnarray}\label{HNR}
    H_{\mathrm{FW}} \simeq \, &\beta \left[ mc^2 -  \frac{1}{4}m{\ddot h}_{lm}^{\mathrm{TT}}{x^l}{x^m}
    \right.\notag\\
    &+ \frac{p^2}{2m} \left(1 - \frac{1}{4c^2} {\ddot h}_{lm}^{\mathrm{TT}}x^l x^m \right)\mathbb{1}_4 \notag\\
    &-  \frac{3\hbar}{16mc^2}{\varepsilon ^{jkl}}{\sigma^l}{\ddot h}_{jm}^{\mathrm{TT}}x^m p_k \mathbb{1}_4 \notag\\
    &- \left.\frac{e^2}{r} \left( 1 - \frac{1}{8c^2}{\ddot h}_{lm}^{\mathrm{TT}}{x^l}{x^m} \right)\right]\mathbb{1}_4.
\end{eqnarray}
It's worth noting that the resulting $H_{\mathrm{FW}}$ takes on a block-diagonal structure. The 4-spinor is decoupled into two irreducible 2-spinor i.e. $\Psi = (\psi_{\mathrm{A}}, \psi_{\mathrm{B}})$, where $\psi_{\mathrm{A}}$ and $\psi_{\mathrm{B}}$ correspond to the particle and antiparticle, respectively.

In the non-relativistic regime, where the energy $E$ of the Dirac particles is much smaller than the energy gap $2mc^2$ between positive- and negative-energy states (i.e., $E \ll 2mc^2$), a single-particle description of the Dirac theory is valid, and the wave function can be interpreted probabilistically. The wave function $\Psi$ satisfies 
\begin{eqnarray}
   \left( {\begin{array}{*{20}{c}}
{{H_A}}&0\\
0&{{H_B}}
\end{array}} \right)\left( {\begin{array}{*{20}{c}}
\psi _A\\
\psi _B
\end{array}} \right) = i\hbar \frac{\partial }{{\partial t}}\left( {\begin{array}{*{20}{c}}
\psi _A\\
\psi _B
\end{array}} \right)
\end{eqnarray}
where $H_A$ and $H_B$ are the positive- and negative-energy parts of the Hamiltonian, respectively. 

The non-relativistic Schr\"{o}dinger equation describes the dynamics of the particle as $H_{\mathrm{NR}} \psi_{\mathrm{A}} = E_{\mathrm{NR}}\psi_{\mathrm{A}}$, with $E_{\mathrm{NR}} = E - mc^2$, and $H_{\mathrm{NR}} \equiv H_A - mc^2$. According to \eqref{HNR}, we obtain
\begin{eqnarray}
   H_{\mathrm{NR}} = H_{\mathrm{Bohr}} + H_{\mathrm{GW}},
\end{eqnarray}
with
\begin{eqnarray}\label{H35}
    H_{\mathrm{Bohr}} = \frac{{{p^2}}}{{2m}} - \frac{{{e^2}}}{r}
\end{eqnarray}
and
\begin{eqnarray}\label{H36}
     H_{\mathrm{GW}} =  - \frac{1}{4}m\ddot h_{lm}^{\mathrm{TT}} x^l x^m - \frac{1}{4c^2}\frac{p^2}{2m}\ddot h_{lm}^{\mathrm{TT}} x^l x^m \notag\\
     + \frac{e^2}{r}\frac{1}{8c^2}\ddot h_{lm}^{\mathrm{TT}} x^l x^m - \frac{3\hbar }{16mc^2}{\varepsilon ^{jkl}} \sigma _l \ddot h_{jm}^{\mathrm{TT}} x^m p_k.
\end{eqnarray}

Based on \eqref{H35} and \eqref{H36}, considering that $p^2/2m \sim e^2/2r \sim (1/2)mc^2\alpha^2$, and $x^i \sim a$, where $\alpha$ is the fine structure constant and $a$ is the Bohr radius, it becomes evident that the first term dominates over the other terms by a factor of $\sim \alpha^{-2}$.

\section{Energy shifts induced by the gravitational waves}\label{secIV}
Our attention is directed towards the computation of energy shifts in the hydrogen atom arising from the predominant influence of GWs, specifically the tidal interaction. This analysis pertains to scenarios where the GW travels through the atom along the z-axis, and  $\ddot h^{\mathrm{TT}}_{ij} = -\omega_{\mathrm{GW}}^2h^{\mathrm{TT}}_{ij}$, with $\omega_{\mathrm{GW}}$ denoting the frequency of the wave. Under these conditions, the interaction Hamiltonian takes the form
\begin{eqnarray}\label{GWH}
    H_{\mathrm{GW}} = \frac{1}{4}m\omega_{\mathrm{GW}}^2h_+(x^2 - y^2).
\end{eqnarray}

The energy shifts can be computed using first-order Perturbation Theory. In our analysis, we assume that the GW interaction term $H_{\mathrm{GW}}$ dominates over the fine structure term, leading to its negligible contribution. Consequently, we treat $H_{\mathrm{GW}}$, given by \eqref{GWH}, as a small perturbation to $H_{\mathrm{Bohr}}$. This perturbative approach is valid when the strength of the GW and its frequency are appropriately chosen.

When computing the energy shifts induced by GWs, it is often necessary to evaluate matrix elements of the form $\langle {n\ell' m'}|x^2 - y^2|{n \ell m}\rangle$. It is useful to know certain properties or selection rules that apply to these matrix elements. We find that 
\begin{eqnarray}
     \left\langle {n'\ell 'm'} \right|{x^2} - {y^2}\left| {n\ell m} \right\rangle  = 0
\end{eqnarray}
unless $\Delta \ell  = ~0,\,\pm2$ and $\Delta m = \pm 2$ (See Appendix B for derivation).

For the ground state ($n=1$), it is nondegenerate, so the energy shift induced by the GW is 
\begin{eqnarray}
    \delta E^{\mathrm{GW}}_{g} = \left\langle {100} \right| H_{\mathrm{GW}} \left| 100 \right\rangle  = 0.
\end{eqnarray}
This means that the ground state is unaffected by the GW perturbation.

For the $n=2$ level, there exists a 4-fold degeneracy. The energy shifts can be determined by finding the eigenvalues of the perturbation matrix with elements $\langle {2\ell'm'} | H_{\mathrm{GW}}| 2\ell m \rangle$:
\begin{eqnarray}
     \left( {\begin{array}{*{20}{c}}
0 &0&0&0\\
0&0 &0&\kappa \\
0&0&0&0\\
0&\kappa &0&0 
\end{array}} \right),
\end{eqnarray}
where $\kappa  \equiv 3ma^2\omega _{\mathrm{GW}}^2h_+(t)$. The energy shifts are $\delta E_{\mathrm{GW}} = 0,\,\pm \kappa$. Consequently, the total energies for each state can be expressed as
\begin{eqnarray}
    E_{\mathrm{I},\mathrm{II}} = E_2 ,\quad E_{\mathrm{III}} = E_2  - \kappa ,\quad E_{\mathrm{IV}}^0 = E_2  + \kappa.
\end{eqnarray}
Their associate eigenstates are
\begin{eqnarray}
    \psi _{\mathrm{I}} &= {\psi _{200}}\\
\psi _{\mathrm{II}} &= {\psi _{210}}\\
\psi _{\mathrm{III}} &= \frac{1}{{\sqrt 2 }}\left( {{\psi _{211}} - {\psi _{21 - 1}}} \right)\\
\psi _{\mathrm{IV}} &= \frac{1}{{\sqrt 2 }}\left( {{\psi _{211}} + {\psi _{21 - 1}}} \right).
\end{eqnarray}

The perturbed light absorption/emission spectra due to the GW (transition from $n=2$ to $n=1$) result in three equally spaced lines corresponding to the photon frequency ratio $ \displaystyle \nu/ \nu_{{\mathrm{Ly}}\alpha } =1,\, 1 \pm 8a^2\omega_{{\mathrm{GW}}}^2h_+/\alpha ^2c^2$, where $\nu_{\mathrm{Ly}\alpha} = (3/8)\alpha ^2mc^2/h$. The transitions probability from $\psi_{\mathrm{II}},\,\psi_{\mathrm{III}},\,\psi_{\mathrm{IV}}$ to the $n=1$ level are dominated by dipole interaction, similar to ordinary Ly$\alpha$ transitions, so the transition could occur when the atom absorbs or emits a photon. However, the transition between $\psi_{\mathrm{III}}$ and $\psi_{\mathrm{IV}}$ is dipole forbidden; the dominated transition probability comes from a non-dipole interaction. This transition is characterized by an alteration of the electron's magnetic quantum number by $\pm 2$, which can be associated with the occurrence of a 2-photon transition \cite{Labzowsky_2005,2S-1S}.

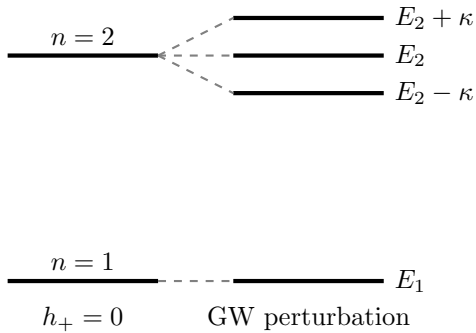
\begin{figure}[h]
    \centering
\tikzset{
    level/.style = {
        ultra thick, black,
    },
    connect/.style = { dashed, gray, thick},
    label/.style = {
        mathrm width=2cm
    }
}
\begin{tikzpicture}
    \draw[level] (0,0) -- node[above] {$n=1$} (2,0);
    \draw[level] (0,3) -- node[above] {$n=2$} (2,3);
    \draw[connect] (2,0) --  (3,0);
    \draw[level] (3,0) -- (5,0) node[right]{$E_1$};
    \draw[level] (3,3) -- (5,3) node[right]{$E_2$};

    \draw[level] (3,2.5) -- (5,2.5) node[right]{$E_2 - \kappa$};
    \draw[level] (3,3.5) -- (5,3.5) node[right]{$E_2 + \kappa$};
    \draw[connect] (2,3) -- (3,3) ;
    \draw[connect] (2,3) -- (3,2.5);
    \draw[connect] (2,3) -- (3,3.5);
    \node at (1,-0.5) {$h_+ = 0$};
    \node at (4,-0.5) {GW perturbation};
\end{tikzpicture}
    \caption{Splitting of $n=2$ states of hydrogen caused by a GW.}
    \label{fig:enter-label}
\end{figure}

\section{Discussion}\label{secV}

In the presence of a GW, an atom's energy levels undergo shifts when compared to those in flat spacetime.  The order of energy shifts induced by GW tidal interactions, as described by \eqref{GWH}, can be estimated. The coordinates $x$ and $y$ are in the vicinity of the atomic radius, meaning that $x \sim y \sim r$, where $r = n^2a$, with $n$ representing the principal quantum number and $a$ denoting the Bohr radius. Consequently, the magnitude of these energy shifts can be expressed as
\begin{eqnarray}
\delta E_{\mathrm{GW}} \propto mh_+\omega _{{\mathrm{GW}}}^2{n^4}{a^2}.
\end{eqnarray}
To provide a perspective on the magnitude of these energy shifts, we can compare them to the hydrogen ground-state hyperfine splitting, denoted as $\delta E_{\mathrm{hf}} \sim (m_e/m_p)\alpha^4mc^2$, which is well-measured in radio astronomy \cite{pritchard2012}. In this expression, $\alpha$ represents the fine-structure constant, while $m_e$ and $m_p$ correspond to the masses of the electron and proton, respectively.

\begin{eqnarray}
\frac{\delta E_{\mathrm{GW}}}{\delta E_{\mathrm{hf}}} \sim \cos(2\theta){n^4}\omega_{\mathrm{GW}}^2h_+ \times 10^{-26} \mathrm{s}^2,
\end{eqnarray}
The $n^4$ dependence indicates a strong impact of GW tidal interactions on atoms in highly excited states, commonly known as Rydberg atoms. Remarkably, Rydberg atoms have been observed in a wide range of astrophysical environments, including interstellar and circumstellar gas clouds \cite{reifenstein1970,gnedin2009}. Furthermore, the dependence of the energy shift on the amplitude and frequency of GWs indicates that the most ideal observable spectrum would come from Rydberg atoms near high-frequency GW sources, exceeding approximately $10\;\mathrm{kHz}$. Such sources are expected to be exotic objects \cite{aggarwal2021}.

The energy splitting induced by GWs exhibits distinct characteristics that differentiate it from other effects, such as Zeeman shifts, Stark shifts, etc., in various atomic states. The interaction between an electron and GW yields an uncommon selection rule, which cannot be achieved through dipole transition. As mentioned above, the transition between states such as $\psi_{\mathrm{III}}$ and $\psi_{\mathrm{IV}}$ requires absorption/emission of two photons. Note that this is the result of the quadrupole term \eqref{GWH} manifesting itself at the level of the first-order correction. Therefore, in the presence of GW, one should expect a higher rate of two-photon transitions. Although this provides a unique indirect observation channel for GWs, the detection of such a transition would be challenging. 

\section{Conclusion}
\label{secVI}

Our investigation into the interaction between a hydrogen atom and GWs provides valuable insights into the behavior of non-relativistic particles in the presence of these dynamic spacetime phenomena. By employing the modified Foldy-Wouthuysen transformation, we successfully derive the Hamiltonian, elucidate selection rules, and quantify the energy shifts induced by GWs. The results obtained enhance our understanding of the effects of GWs on the dynamics of the hydrogen atom and lay the foundation for exploring the broader implications of GWs in quantum systems.

\ack
We sincerely thank Prof. Mairi Sakellariadou for her valuable comments and suggestions. We also acknowledge the support of NSRF through the Program Management Unit for Human Resources and Institutional Development, Research, and Innovation (grant number B05F650024).

\appendix
\section{Computing the relevant commutators}
The various terms in \eqref{HH} are computed as follows: 
\begin{eqnarray}
    i[S,H] = i[S,H_0] + i[S,H_I],
\end{eqnarray}
where
\begin{eqnarray}
i[S,H_0] =  - c{\alpha ^k}{p_k} + \frac{p^2}{m}\beta  + \frac{i\hbar e^2}{2mc}\beta {\alpha ^k}\partial _k\left( \frac{1}{r} \right) \\
i[S,H_I] =  - \frac{{i\hbar }}{{4c}}\ddot h_{mk}^{{\mathrm{TT}}}{x^m}{\alpha ^k} + \frac{1}{{4c}}\ddot h_{lm}^{{\mathrm{TT}}}{x^l}{x^m}{p_k}{\alpha ^k} \notag\\
+ \frac{{i\hbar }}{{4m}}\beta {\delta ^{jk}}\ddot h_{jm}^{{\mathrm{TT}}}{x^m}{p_k} - \frac{\hbar }{{4m}}\beta {\varepsilon ^{jkl}}{\sigma ^l}\ddot h_{jm}^{{\mathrm{TT}}}{x^m}{p_k} \notag\\
- \frac{1}{{4m}}\beta \ddot h_{lm}^{{\mathrm{TT}}}{x^l}{x^m}{p^2} + \frac{{i{\hbar ^2}}}{{8m}}\beta {\varepsilon ^{jkl}}{\sigma ^l}\ddot h_{jk}^{{\mathrm{TT}}} \notag\\
 + \frac{{i\hbar }}{{4m}}\beta {\delta ^{jk}}\ddot h_{mk}^{{\mathrm{TT}}}{x^m}{p_j} 
 - \frac{{i{e^2}\hbar }}{{16mc}}\beta {\alpha ^k}{\partial _k}\left( {\frac{1}{r}} \right)\ddot h_{lm}^{{\mathrm{TT}}}{x^l}{x^m}\notag\\
 - \frac{{i{e^2}\hbar }}{{8mc}}\beta {\alpha ^k}\ddot h_{km}^{{\mathrm{TT}}}\frac{{{x^m}}}{r},
\end{eqnarray}
and the last commutator is
\begin{eqnarray}
\frac{i^2}{2}[S,[S,\beta mc^2(1 - \frac{1}{4c^2}\ddot h_{lm}^{\mathrm{TT}}{x^l}{x^m})]] \notag\\
=  - \frac{p^2}{2m}\beta  + \frac{1}{8mc^2}{p^2}\ddot h_{lm}^{\mathrm{TT}}{x^l}{x^m}\beta \notag\\
+ \frac{\hbar }{{16m{c^2}}}{\varepsilon ^{jkl}}{\sigma ^l}\ddot h_{jm}^{{\mathrm{TT}}}{x^m}{p_k}\beta \notag \\ 
 - \frac{{i{\hbar ^2}}}{{16m{c^2}}} {\sigma ^l}{\varepsilon ^{jkl}}\ddot h_{jk}^{{\mathrm{TT}}}\beta - \frac{{3i\hbar }}{{16m{c^2}}} {\delta ^{jk}}\ddot h_{mj}^{\mathrm{TT}}{x^m}{p_k}\beta.
\end{eqnarray}

\section{The selection rule for $\langle n'\ell'm'|x^2 - y^2|n \ell m\rangle$}\label{secIVA}
The Wigner-Eckart theorem provides a means of calculating the matrix elements for rank-$k$ spherical tensor operators, denoted as $T^k_q$, with respect to angular momentum eigenstates. It is expressed as follows:
\begin{eqnarray}\label{Wigner}
\langle n'\ell 'm'|T_q^k | n\ell m \rangle  = \langle {\ell k;mq}| \ell' m' \rangle \left\langle {\left. {n'\ell'} \right\|} \right.T^k\left\| {\left. {n \ell} \right\rangle } \right.,
\end{eqnarray}
where, $\langle \ell k;mq| \ell' m' \rangle$ represents the Clebsch-Gordan coefficient associated with the combination of angular momenta $\ell$ and $k$ to yield $\ell'$. The double-bar matrix element, $\left\langle {\left. {n'\ell'} \right|} \right.T^k\left| {\left. {n \ell} \right\rangle } \right.$, is referred to as the reduced matrix element, a factor independent of $m$, $m'$, and $q$.

In terms of spherical harmonics, 
\begin{eqnarray}
    \left\langle {n'\ell 'm'} \right|x^2 - y^2\left| n\ell m \right\rangle  \notag\\
    = \sqrt {\frac{{8\pi }}{{15}}} \left\langle {n'\ell '} \right|r^2\left| {n\ell } \right\rangle \left\langle {\ell 'm'} \right|Y_2^{ - 2} + Y_2^2\left| \ell m \right\rangle,   
\end{eqnarray}
where
\begin{eqnarray}
    \left\langle {n\ell '} \right| r^2 \left| {n\ell } \right\rangle  = \int {R_{n\ell '}^*(r) r^2 {R_{n\ell }}(r){r^2}dr}.
\end{eqnarray}
According to the Wigner-Eckart theorem, we have 
\begin{eqnarray}
   \left\langle {\ell 'm'} \right|Y_2^{ \pm 2}\left| {\ell m} \right\rangle  = \left\langle {{\ell 2; \pm 2m}}
 \mathrel{\left | {\vphantom {{\ell 2; \pm 2m} {\ell 'm'}}}
 \right. \kern-\nulldelimiterspace}
 {{\ell 'm'}} \right\rangle \left\langle {\left. {\ell '} \right\|} \right.{Y_2}\left\| {\left. \ell  \right\rangle } \right..
\end{eqnarray}
So, we have
\begin{eqnarray}
    \left\langle {n'\ell 'm'} \right|{x^2} - {y^2}\left| {n\ell m} \right\rangle  \notag\\
    = \sqrt {\frac{{8\pi }}{{15}}} \left\langle {n'\ell '} \right|{r^2}\left| {n\ell } \right\rangle \times \\
    \left\langle {\left. {\ell '} \right\|} \right.{Y_2}\left\| {\left. \ell  \right\rangle } \right.\left[ {\left\langle {{\ell 2; - 2m}}
 \mathrel{\left | {\vphantom {{\ell 2; - 2m} {\ell 'm'}}}
 \right. \kern-\nulldelimiterspace}
 {{\ell 'm'}} \right\rangle  + \left\langle {{\ell 2;2m}}
 \mathrel{\left | {\vphantom {{\ell 2;2m} {\ell 'm'}}}
 \right. \kern-\nulldelimiterspace}
 {{\ell 'm'}} \right\rangle } \right].
\end{eqnarray}
The rules of addition of angular momenta require that 
\begin{eqnarray}\label{S1}
\left\langle \ell 2; \pm 2m | \ell 'm'\right\rangle  = 0,
\end{eqnarray}
unless $\Delta m =  \pm 2$ and $\Delta \ell =  0, \pm 1, \pm 2$.
Consequently, under rotational symmetry,
\begin{eqnarray}
    \left\langle {n'\ell 'm'} \right|{x^2} - {y^2}\left| {n\ell m} \right\rangle = 0
\end{eqnarray}
unless $\Delta m =  \pm 2$ and $\Delta \ell =  0, \pm 1, \pm 2$.

Additionally, let's explore the parity invariance of
\begin{eqnarray}
  \left\langle {\ell 'm'} \right|Y_2^{ - 2} + Y_2^2\left| \ell m \right\rangle = \notag\\
     \left[ \begin{array}{l}
   \displaystyle  \iint{{{[Y_{\ell '}^(\theta ,\phi )]}^*}Y_2^{ - 2}(\theta ,\phi )Y_\ell ^m(\theta ,\phi )d\Omega }
\\
\displaystyle + \iint {{{[Y_{\ell '}^{m'}(\theta ,\phi )]}^*}Y_2^2(\theta ,\phi )Y_\ell ^m(\theta ,\phi )d\Omega }
\end{array} \right].
\end{eqnarray}
Utilizing the property $[Y_\ell ^m(\theta ,\phi )]^* = ( - 1)^m Y_\ell ^{- m}(\theta ,\phi )$ and $Y_\ell ^m(\pi  - \theta ,\phi  + \pi ) = {( - 1)^\ell }Y_\ell ^m(\theta ,\phi )$, we can deduce that
\begin{eqnarray}\label{S3}
    \left\langle n \ell 'm' \right| x^2 - y^2 \left| n \ell m \right\rangle  = (- 1)^{\ell ' + \ell  + 2} \notag\\
   \times \left\langle n \ell 'm' \right|x^2-y^2\left| n \ell m \right\rangle.
\end{eqnarray}
This leads to the conclusion that the matrix element vanishes unless  $\Delta \ell = 0,, \pm 2,\ldots $.

By combining \eqref{S1} and \eqref{S3}, we arrive at the condition
\begin{eqnarray}
     \left\langle {n'\ell 'm'} \right|{x^2} - {y^2}\left| {n\ell m} \right\rangle  = 0,
\end{eqnarray}
unless  $\Delta \ell  = ~0,\,\pm2$ and $\Delta m = \pm 2$.

\section*{References}
\bibliography{GW}
\bibliographystyle{unsrt}
\end{CJK*}
\end{document}